# 5G-CORNET: Platform as a Service

Vuk Marojevic, Shem Kikamaze, Randall Nealy, Carl Dietrich
Wireless@Virginia Tech, Bradley Dept. Electrical and Computer Engineering
Virginia Tech, Blacksburg, Virginia USA
{maroje, shemk, rnealy, cdietric}@vt.edu

*Abstract*—Practical testing of the latest wireless communications standards requires the availability of flexible radio frequency hardware, networking and computing resources. We are providing a Cloud-based infrastructure which offers the necessary resources to carry out tests of the latest 5G standards. The testbed provides a Cloud-based Infrastructure as a Service. The research community can access hardware and software resources through a virtual platform that enables isolation and customization of experiments. In other words, researchers have control over the preferred experimental architecture and can run concurrent experiments on the same testbed. This paper introduces the resources that can be used to develop 5G testbeds and experiments.

## I. INTRODUCTION

Wireless networks must meet rapidly increasing demands for bandwidth, mobility, and reliability. Cognitive radio (CR) combines cognitively inspired learning and decision making with reconfigurable software-defined radios (SDRs) to address this need and maximize quality of service and communications capacity within physical constraints, such as limited spectrum availability or battery-supplied energy. This area of research is extremely timely; once mature, CR technology will enable dynamic spectrum sharing, including the sharing of 1000 MHz of spectrum currently used exclusively by the Federal Government. This was recommended to the President as a way to stimulate economic growth by the President's Council of Advisors on Science and Technology (PCAST) [1]. The PCAST report indicates that dynamic spectrum sharing would increase capacity of this Federal spectrum by a factor of 1000, as a conservative estimate.

Experimentation and validation on large scale physical systems are required to move CR research forward. We seek to enable the next leap in research by providing a unique resource to a wide research community. Experience with modern large-scale university testbeds will significantly reduce the risk associated with developing even greater testbeds (e.g., an urban "Test City in a major U.S. city along with a Mobile Test Service that can relocate to urban, rural, and Federal facilities, to support rapid experimentation in spectrum management technology and practice," as recommended by PCAST [1].)

Whereas CR and SDR research has been abundant over the past 20 years [2]—a Google Scholar search for "cognitive radio" returned more than 1.3M results—, the advent of the Internet of Things, industry 4.0, intelligent transportation and unmanned aerial systems, augmented reality, and so forth, bring along new challenges for wireless communications and its integration with high-performance, distributed and Cloud computing.

Our mission is to (1) create a professional user-experience for a broad community of users to transition research beyond current limitations, and to (2) educate the next generation of professionals on emerging and new IT and communications technology.

This paper presents our new cognitive radio network (CORNET) testbed, 5G-CORNET, and the research it enables. Section II discusses prior work. Section III presents the testbed hardware and software infrastructure. Section IV briefly discusses some of the enabled research before concluding the paper in Section V.

## II. RELATED WORK

Testbeds have played an important role in the development of networking technologies since the first computer networks were created. In planning our testbed, we were mindful of academic CR testbeds deployed in the past few years. The WINLAB ORBIT testbed at Rutgers University is a large-scale wireless network testbed with over four hundred nodes. Most of these nodes consist of a PC equipped with 802.11 a/b/g network cards [3]. The original testbed was supplemented with flexible radio frequency (RF) platforms to exploit physical layer adaptations and is continuously upgraded with modern hardware and software. ORBIT is part of the GENI network [4].

The wireless networking research group at the University of California, Riverside has also deployed a testbed consisting of fifty-eight 802.11 wireless nodes with additional multiple input multiple output (MIMO) cards, fifteen laptops connected to Universal Software Radio Peripherals (USRPs), and six laptops connected to the Rice University WARP radios [5]. Like ORBIT, the majority of the nodes in this testbed are targeted to upper layer network protocol research. However, this testbed includes both the WARP boards and the USRP, which have flexible and powerful processors onboard. The Emulab research facility [6] at the University of Utah is a large testbed of approximately 275 nodes with heterogeneous hardware characteristics.

Virginia Tech's (VT's) CORNET testbed has been designed and developed to extend and augment the wireless testbed concepts implemented at other universities. CORNET benefits the academic community by providing a stable and flexible platform for research and education [7], while advancing the level of relevance testbeds can bring to CR applications. Our approach to testbed design is unique in a number of key ways:

Primarily, we aim to provide a high degree of flexibility in terms of radio waveform implementation and adaptive capabilities. We found that most existing testbeds make use of radio hardware tailored to a single wireless technology—be it Wi-Fi, Bluetooth, or cellular communications system—or combine equipment of similar technologies. Since most CR applications require radios with adaptive link capabilities that these protocols do not implement, these testbeds impose important restrictions for achieving our research and educational goals.

The desire for a real-world testing environment arose from talking to researchers in the area, who routinely commented that their prototype applications performed significantly worse once they took their work outside the laboratory. Researchers have the option to use adjacent, line-of-sight nodes in combination with nodes located in different parts of the building, or on different floors to emulate a number of meaningful and realistic wireless networking scenarios. This combination of non-ideal channel conditions represents a philosophical departure from previous testbed designs and offers a relevant CR prototyping environment.

Among existing CR testbeds, CORNET is unique in terms of the number of deployed SDR devices and multiple licensed bands of operation ranging from 138 to 3600 MHz. The upgraded CORNET testbed described in this paper can be used to define 5G networks or network elements and enables emerging research on spectrum and network sharing. 5G-CORNET can be integrated with two other VT testbeds—Outdoor-CORNET and LTE-CORENT—which would enable researchers to carry out a variety of realistic experiments not possible with just an indoor testbed.

## III. TESTBED INFRASTRUCTURE

### A. Testbed Overview

5G-CORNET provides wide-band RF hardware (Ettus USRP X310 units and wide-band daughter boards) paired with new high-performance computing cluster nodes. The USRP X310s are the RF front end SDRs, or remote radio heads (RRHs), whereas the Xeon computers act as hosts computers interconnected through a flexible 10 Gigabit Ethernet network switch (Fig. 1).

### B. Software-Defined Radios

The testbed uses a mix of USRP X310s and the older generation USRP2s, which will eventually be phased out. Each X310s has two UBX daughter boards, which have an instantaneous bandwidth of up to 160 MHz with a maximum center frequency of 6 GHz. Each X310 has two receivers and two transceivers, which are can be utilized for MIMO operation.

The antenna array on each USRP serves multiple purposes depending on the desired mode. The array has an element connected to each USRP transmit/receive port on the node (total of 4 ports). Monitoring functions will be enabled through the receive port of the USRP daughterboard using a RF coupler. With this setup, MIMO, diversity and beamforming experiments are possible. The antenna ports can also be used independently. The antennas are installed in or above the drop-ceiling adjacent to the node. The new antennas are placed clear of obstructions. An analog RF channel emulator can be used for accessing frequency bands without the need for a license [8].

### C. Computing Cluster

The computing cluster currently consists of 10 new rackmount computers. Each computer has dual 3.0 GHz 12-Core Xeon processors with 128 GB RAM, expandable to 1.54 TB. Resources for the different users will be divided based on the available computing resources. After gaining more data on system utilization, various upgrades will be made to accommodate future needs.

### D. Network Infrastructure

The system is interconnected through the 10 Gigabit Ethernet network switch with 96 ports. The switch introduces a minimal latency of 550 ns, which ensures that there is low latency between the various computers in the cluster. The routing between different ports is controlled through a management node. Any USRP is accessible by any computer through the switch. This also ensures that additional USRPs and computers can be added to expand the network and accommodate future research needs.

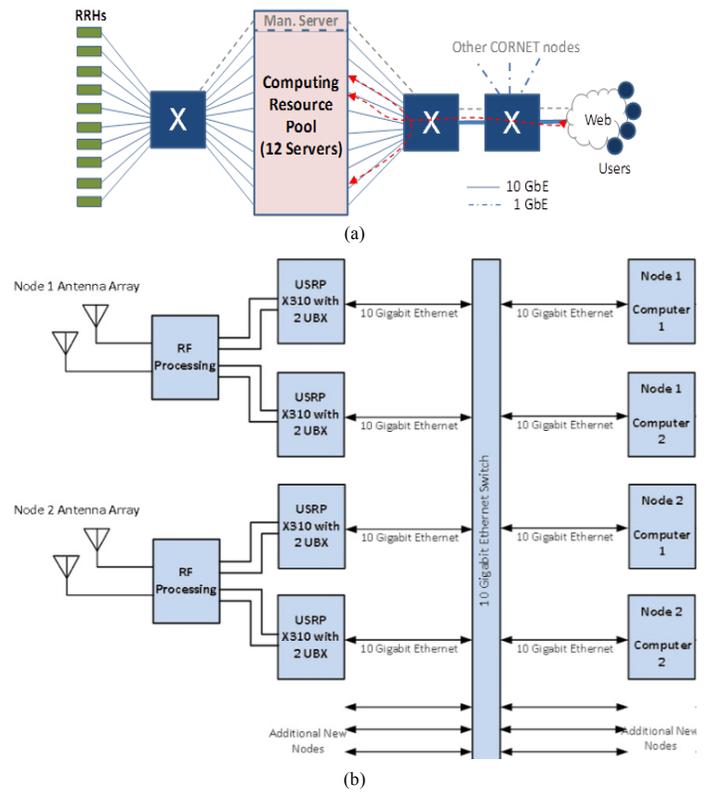

Figure 1: 5G-CORNET infrastructure (a) and RF plus computing network (b).

## IV. RESOURCE MANAGEMENT

In order to support the use of SDR cluster computing in the testbed, software tools will control the assignment of processing tasks to computing nodes. Different layers of research management are supported by 5G-CORNET.

### A. Public Web Portal

The first software development phase focuses on building the public web portal. It is through this interface that most users will interact with 5G-CORNET to schedule time on the network to perform experiments and analyses. At the heart of the public portal is the graphical user interface (GUI). The GUI will be a publicly available Web Site. Development of this Web Site will utilize an Apache, PHP, MySQL technology stack on the server side and may utilize JavaScript, Flash, Java, and HTML5 in the browser. The GUI will interact with CORNET via an application programming interface (API). The development of a basic API framework will enable connectivity between the GUI and existing CORNET capabilities. An example screenshot of the proposed web portal is shown in Fig. 2.

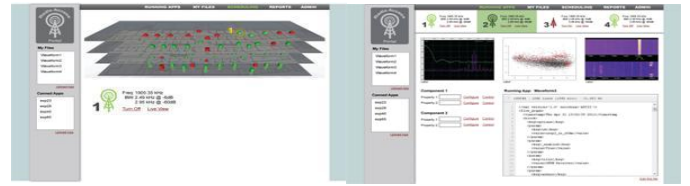

Figure 2: Proposed GUI access to 5G-CORNET: floorplan and radio statistics.

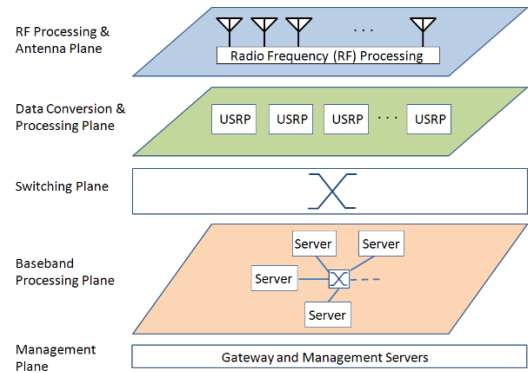

Figure 3: 5G-CORNET: layered architecture.

The main GUI navigation will allow access to each new component as it is implemented. This will allow users not familiar with the underlying architecture of 5G-CORNET to easily access the network to schedule experiment time on various nodes, view ongoing test results, and easily capture information about each experiment.

*B. Clustering for Distributed Computing*

Cluster computing enhances the computational ability of a single node computer by combining the resources of two or more computers to increase the data processing ability. SDR applications often have components that may be processed either in parallel or with a pipeline architecture. The 10 Gigabit switching network of the new testbed allows for efficient passing of data between node computers and between the different processing planes, as illustrated in Fig. 3. The high-speed and flexible Ethernet network support various clustering configurations.

*C. Resource Virtualization*

Resource virtualization includes virtual machines (VMs) for isolating access to shared resources and spectrum virtualization, which extends RF reuse. VMs are used for isolating access to shared computing resources to allow users to run independent software on a single node computer even when the software has incompatible requirements. VMs, which can implement virtual network functions or entire communications systems [9], can be operated in combination with RF resource virtualization to support multiple users or virtual wireless network on the same physical network.

Spectrum virtualization facilitates accessing RF resources by multiple users simultaneously. Baseband signals generated in software can be offset in frequency, then summed and passed to the RF hardware for transmission as a spectrum block. Conversely, a spectrum block is processed and divided into segments for independent demodulation. This technique readily makes use of cluster processing. With a nominal bandwidth of 160 MHz per USRP it is possible to process a total bandwidth of at least 320 MHz at each dual-USRP node. This is because we have two UBX daughter boards per node, but the bottleneck is the 10 Gigabit network. A 320 MHz signal could include a combination of several LTE base stations along with legacy communications systems in the same physical dual-node.

*D. Enhanced Scheduler*

The scheduler will allow 5G-CORNET users to remotely verify the availability of nodes and potential time frames to request access accordingly. As users make requests, the scheduler makes tentative reservations within the system database. Some reservations may be immediately confirmed, whereas others may require the system administrator to review the request prior to approval. The scheduler will also keep data about scheduled and actual usage of resources for real-time control and post-processing. As a part of the assessment process there will be a required post-usage survey generated by the Scheduler and distributed to the user. Committee schedule reviews and other automated methods are currently considered to ensure fair access to resources and yet allow large experiments.

The scheduler allocates computing, network and radio resources to the end user (Fig. 4). There will be preconfigured nodes with preconfigured software that the end-user can use. More advanced users with specific needs will be able to select features, such as:

a. Computing resources:
   - RAM,
   - Hard Drive ROM,
   - Duration of life for the VM,
   - Amount of CPU Threads and Cores and any software needed for experimentation, such as CRTS [10], srsLTE, GNURadio, LiquidDSP, or VT's open source SAS [11].

b. Radio resources:
   - The number of USRPs connected to a virtual node,
   - The frequency channels to be utilized from a range of available experimental licensed bands,
   - Instantaneous bandwidth for each radio,
   - The ability to choose between over the air transmission or transmission through the RF channel emulator.

c. Network resources:
   - Network resources will be based on the current load and the requested bandwidth for each virtual node.

To accomplish this, the scheduler has control over both the Ethernet switch required to route traffic and the computing cluster to allocate computing resources.

*E. Data Manager*

The 5G-CORNET testbed is capable of producing vast quantities of data. An example would be using multiple nodes to produce spectra and (through the antenna arrays) directional data simultaneously. In such case, data would have a frequency component, a directional component for each frequency and a location component for each physical node used. All of those components would also be referenced to time. The data manager needs to store the available data in a consistent format for processing. Archiving and data storage will be done in a systematic manner along with system and device configurations so that the data will be useful to future uses.

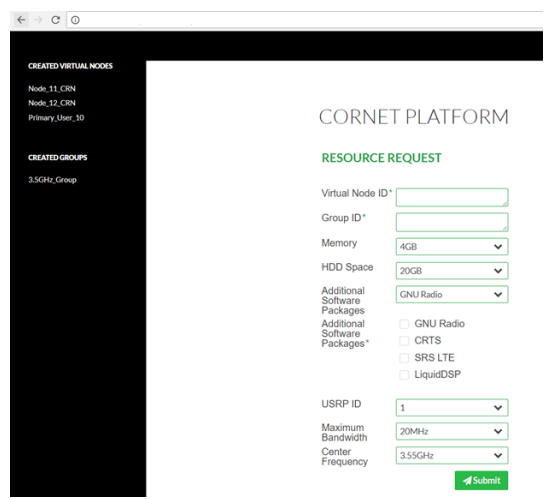

*Figure 4: Web interface for requesting testbed resources.*

## V. ENABLED RESEARCH

The new testbed integrates powerful SDRs with a computing cluster and uses virtualization technology to schedule users, waveforms, and networks. LTE-CORNET and CRTS are currently being added and integrated into the system. LTE-CORNET provides a platform for testing various LTE scenarios through a channel emulator or over the air [10]. CRTS is an OFDM-based platform for testing performance of different SDRs under various interference conditions [7].

Virginia Tech's open-source spectrum access system (SAS) [8] is being developed and tested on 5G-CORNET. The model for the interaction between the SAS, primary and secondary users can be done through the wireless network or the cabled setup using the channel emulator. The channel emulator can emulate physical location of nodes. This allows reproducible testing using diverse radio environment and emulated geographical regions.

Fig. 5a illustrates the Ethernet switch at the top, the computing cluster and five SDRs that are connected to the analog RF channel emulator RFnest [12] on the bottom. RFnest defines RF attenuation based on channel models or empirical data to emulate a desired theoretical or real-world scenario (Fig. 5c). Researchers have access to RFnest in order to create the propagation environment that suits their needs. Along with up to eight physical radios, additional simulated or virtual radios can be added in software to create a variety of RF conditions. Fig. 5d depicts the internal network for our SAS testbed using 5G-CORNET. The internal virtual network and the external Juniper network switch enable routing I/Q data or control messages between VMs and USRPs or other hardware, such as the RFnest. Fig. 5b shows one of the dual-node RRHs that are deployed underneath the ceiling for over the air experiments.

## I. CONCLUSIONS

The upgraded CORNET testbed provides a PaaS for researchers who are interested in creating a 5G testbed with the resources described in this paper. The available RF, computing and networking resources can be used to build Cloud-based wireless networks with powerful and flexible SDRs and a centralized computing cluster. The collocated pairs of SDRs facilitate implementing MIMO systems or monitoring spectrum activities. The channel emulator allows defining a variety of radio environments—static as well as dynamic. The flexibility of the platform allows building custom testbeds and defining reproducible experiments.

## ACKNOWLEDGEMENT

This work was in part supported by NSF through award numbers CNS-1629935, CNS-1642873, and CNS-1564148.

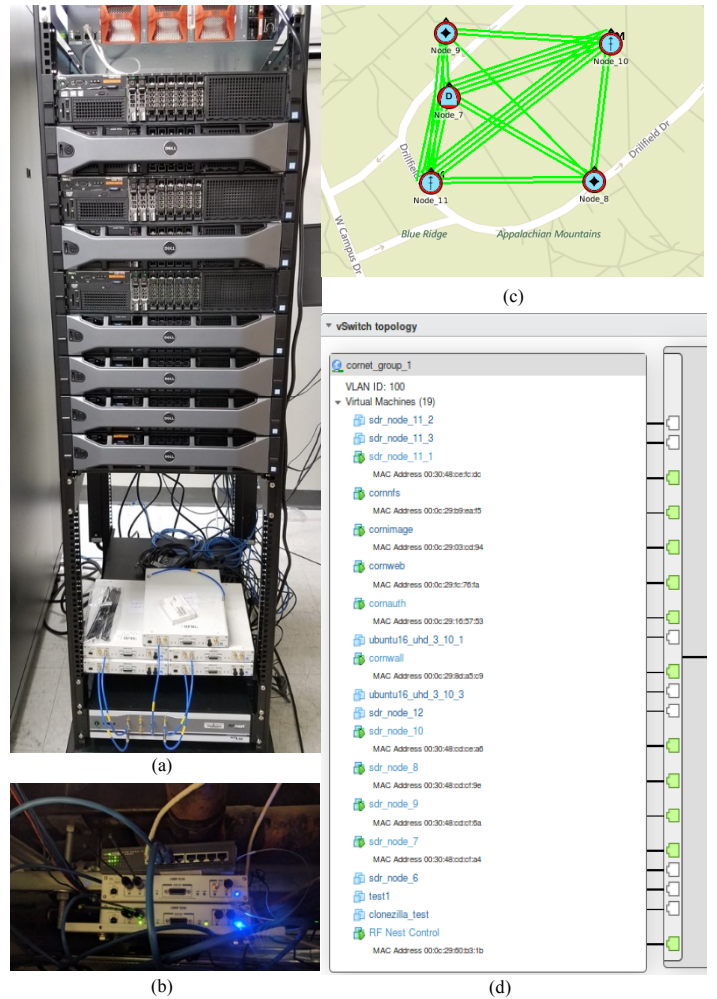

*Figure 5: 5G-CORNET testbed: Ethernet switch (backside), nine-node computing cluster, five local USRPs and RF channel emulator (a, top to bottom), dual node RRH (b), RFview software defining emulated radio locations and RF channel conditions (c), and computing cluster internal network showing 19 VMs that act as management or SDR computing nodes (d).*